\documentclass{ijuc}
\usepackage{graphicx,amsmath,amssymb}
\newtheorem{theorem}{Theorem}
\newtheorem{proposition}{Proposition}
\newtheorem{corollary}{Corollary}

\begin{document}

\title{Enumeration of number-conserving cellular automata rules
with two inputs}

\author{Henryk Fuk\'s\email{hfuks@brocku.ca}
\and
Kate Sullivan
}

\institute{Department of Mathematics, Brock University\\
St. Catharines, Ontario L2S 3A1, Canada}

\maketitle

\begin{abstract}
We show that there exists a one-to-one correspondence between
the set of number-conserving cellular automata (CA) with $q$  inputs 
and the set of balanced sequences with $q$ terms. This allows to enumerate
number-conserving CA. We also show that number-conserving 
rules are becoming increasingly rare as the number of states increases.
\end{abstract}


\section{Introduction}
Cellular automata (CA) rules possessing additive invariants have been studied since
early 90's. The simplest form of an additive invariant is the sum of
all site values over a finite lattice with periodic boundary conditions.
Rules with such an invariant are known as number-conserving rules,
and they exhibit many interesting properties. For example,
they can be viewed as  systems of interacting particles
\cite{paper10}. In a finite system, the flux or current of
particles in equilibrium depends only on their density. The graph of the current as a function of density
characterizes many features of the flow, and is therefore called
the fundamental diagram. Fundamental diagrams of number-conserving CA
exhibit intriguing singularities,
which have been investigated extensively \cite{paper8,paper19}.

Although the general conditions for existence of additive invariants
are known \cite{Hattori91}, not much is known about the distribution 
of CA rules possessing invariants in the space of all CA
rules. 
In this paper, we will demonstrate that for the  case of two-input
rules, number-conserving CA are equivalent to so-called 
balanced sequences, and, therefore, can be enumerated. We will also
show that the proportion of number-conserving rules among all CA rules is decreasing
and tends to zero as the number of states increases. 
\section{Definitions}
Let $\mathcal{Q}$ be a finite set of states,
equal to $\{0,1,2,\ldots,q-1\}$. Given  a positive integer $n$, let 
$f: \mathcal{Q}^n\to \mathcal{Q}$ be a local function of one-dimensional cellular 
automaton, also called \emph{CA rule}. The rule $f$
is \textit{number-con\-serving} if, for all cyclic configurations $(x_1,x_2,\ldots, x_L)\in \mathcal{Q}^L$ of length
$L\ge n$, it satisfies
\begin{multline}
f(x_1,x_2,\ldots,x_{n-1},x_n)+f(x_2,x_3,\ldots,x_n,x_{n+1})+\cdots\\
+f(x_L,x_1\ldots,x_{n-2},x_{n-1})=x_1+x_2+\cdots+x_L.
\label{CRf}
\end{multline}
The following characterization of number-conserving rules will be useful in subsequent
considerations. It is a special case of a general result of Hattori and Takesue \cite{Hattori91},
which has been recently generalized and further developed by several other authors \cite{Pivato02,Durand2003,Formenti2003,Moreira03}.

\begin{theorem}A one-dimensional $q$-state $n$-input CA rule $f$ is
number-con\-serving if, and only if, for all
$(x_1,x_2,\ldots,x_n)\in{\mathcal{Q}}^n$, it satisfies
\begin{align} 
&f(x_1,x_2,\ldots,x_n) = x_1 +\notag\\
& \sum_{k=1}^{n-1}\big(
f(\underbrace{0,0,\ldots,0}_k,x_2,x_3,\ldots,x_{n-k+1})
-f(\underbrace{0,0,\ldots,0}_k,x_1,x_2,\ldots,x_{n-k})\big).
\label{NScond}
\end{align}
\end{theorem}

In this paper, we will be interested in enumeration of number-conserving rules.
Let $\Lambda_{n,q}$ be the number of number-conserving rules with $n$ inputs
and $q$ states. The general formula for $\Lambda_{n,q}$ is not currently known,
and finding such formula appears to be rather difficult. Therefore, one could
at first attempt to attack special cases. For $n$-input binary rules, first five terms
of the sequence $\{\Lambda_{n,2}\}_{n=1}^\infty$ have been obtained by a
direct computer-assisted search~\cite{paper8}. These terms are $\{\Lambda_{n,2}\}_{n=1}^{5}=1,2,5,22,428$.
Unfortunately, closed-form expression for this sequence remains unknown.

Since $q=2$ is the smallest number of states yielding non-trivial CA, one could
also consider a somewhat complementary case, namely the smallest $n$ which yields non-trivial rules,
which is obviously $n=2$.
We will therefore  attempt to find $\Lambda_{2,q}$. For two-input rules,
condition (\ref{NScond}) simplifies significantly, becoming
\begin{equation} \label{n2cond}
f(x_1,x_2) = x_1 + f(0,x_2)-f(0,x_1) \,\,\,\text{for all $x_1,x_2 \in \mathcal{Q}$}.
\end{equation}
Obviously, we also require that all outputs of $f$ remain in the set $\mathcal{Q}$, that is,
\begin{equation}\label{bounds}
0 \leq f(x_1,x_2) \leq q-1 \,\,\,\,\,\text{for all $x_1,x_2 \in \mathcal{Q}$}.
\end{equation}

\section{Balanced sequences}
We will now demonstrate that conditions (\ref{n2cond}) and (\ref{bounds}) can be reduced to a set of
double inequalities. First, it will be useful to write conditions (\ref{n2cond}) and (\ref{bounds})  separately
for $x_1>0$ and $x_1=0$ cases. This yields
\begin{align}
&f(x_1,x_2) = x_1 + f(0,x_2)-f(0,x_1), \label{a1} \\
&0 \leq f(x_1,x_2) \leq q-1, \label{a2} \\
&f(0,x_2) =  f(0,x_2)-f(0,0),\label{a3}  \\
&0 \leq f(0,x_2) \leq q-1,\label{a4}  
 \end{align}
 where $x_1\in \mathcal{Q} \setminus \{0\} $ and $x_2 \in \mathcal{Q}$.
Notice that  for $x_2=0$, eq. (\ref{a3}) leads to $f(0,0)=0$.
Taking this into account, (\ref{a3}) becomes $f(0,x_2)=f(0,x_2)$, hence we can treat
$f(0,x)$  for $x\in\mathcal{Q}$ as ``free'' parameters. Let us define $a_1,a_2,\ldots,a_q$ by
$a_{q-i}=f(0,i)$ for all $i\in \mathcal{Q}$.
Using this notation,  eq. (\ref{a1}) becomes
\begin{equation}\label{outputs}
f(i,j)=i+a_{q-j}-a_{q-i} \mbox{\,\, for all $i,j\in \mathcal{Q}$.}
\end{equation}
This means that conditions (\ref{a1}--\ref{a4})
 reduce to
 \begin{align} 
 &a_q=0, \label{singlecond0}\\
 &0 \leq a_k \leq q-1, \label{singleconda} \\
 &0 \leq i+a_{q-j}-a_{q-i} \leq q-1, \label{singlecondb}
 \end{align}
where $k\in\{1,2,\ldots q\}$, and $i,j \in\mathcal{Q}$. In other words, a number-conserving CA rule
is uniquely defined by a sequence of $q$ integers $a_1$,$a_2$, $\ldots$, $a_q$ satisfying 
the condition (\ref{singlecond0}) and
the set of double inequalities
(\ref{singleconda}) and (\ref{singlecondb}) . If we know the sequence $a_1,a_2, \ldots, a_q$, we can determine all outputs of $f$ by using equation (\ref{outputs}). We will say that $a_1,a_2, \ldots, a_q$ is the {\it defining sequence}
of $f$. Table 1 shows examples of number-conserving rules and their defining sequences
for $q=2,3$, and $4$.

In what follows, it will be more convenient to work with slightly transformed version of inequality
(\ref{singlecondb}). Defining $k=q-i$, $l=q-j$, and rewriting  (\ref{singlecondb}) with these
indices one obtains the
following proposition.
\begin{proposition} \label{th2}
Two-input $q$-state CA rule $f$ is number-conserving if and only if for all $i,j \in \mathcal{Q}$ 
\begin{equation}
f(i,j)=i+a_{q-j}-a_{q-i},
\end{equation}
where $a_1,a_2,\ldots, a_q$ is a sequence of integers satisfying
\begin{align} 
&a_q=0 \label{line0} \\
&0 \leq a_k \leq  q-1, \label{line1} \\
&0 \leq q-k + a_l - a_k \leq q-1,\label{line2}
\end{align}
for all $k,l\in \{1,2,\ldots, q\}$.
\end{proposition}
\begin{table}
\begin{center}
\begin{tabular}{lll}
 $q=2$ \hspace{1cm}  &  $q=3$  \hspace{2cm} & $q=4$  \\ 
11\fbox{00} &222111\fbox{000}    & 333322221111\fbox{0000}\\
10\fbox{10} &211211\fbox{100}    & 322232222111\fbox{1000}\\
     &221110\fbox{110}    & 332222112211\fbox{1100}\\
     &210210\fbox{210}    & 321132113211\fbox{2100}\\
     &                    & 331122003311\fbox{2200}\\
     &                    & 323232321010\fbox{1010}\\
     &                    & 333222211110\fbox{1110}\\
     &                    & 322132212110\fbox{2110}\\
     &                    & 332122102210\fbox{2210}\\
     &                    & 321032103210\fbox{3210}
\end{tabular}
\end{center} 
 \caption{Two-input number-conserving rules with two, three, and four
 states.
 Each rule is represented by a sequence of output values $b_{q^2-1},b_{q^2-2},\ldots, b_0$, 
 where $b_{q x+y}=f(x,y)$ for $x,y \in \mathcal{Q}$. In particular, for all $i \in 
 \mathcal{Q}$ one has $b_i=f(0,i)=a_{q-i}$, i.e., in each line the framed suffix
 $b_{q-1} \ldots b_0=a_1 \ldots a_q$ is the defining sequence of the rule.
 }
\end{table}
Conditions (\ref{line1}) and (\ref{line2}) can be simplified even further. First of all, 
(\ref{line1}) and (\ref{line2}) are equivalent to four inequalities
\begin{align} \label{part1}
a_k &\geq 0  \\
a_k &\leq  q-1,\label{part2} \\
a_k &\leq q-k+a_l,\label{part3}\\
a_k &\geq 1-k+a_l,\label{part4}
\end{align}
for all $k,l\in \{1,2,\ldots, q\}$.
Inequality (\ref{part3}) is equivalent to $a_k\leq q-k + \min\{a_l\}$, but $\min\{a_l\}=a_q=0$,
thus we obtain  $a_k\leq q-k$.

When $k=1$, inequality (\ref{part4}) leads to $a_1 \geq a_l$ for all $l \in \{1,2,\ldots, q\}$, which means that
$a_1$ is the largest term of the sequence. Therefore, (\ref{part4}) is equivalent to
$a_1 \geq a_k \geq 1-k+a_1$ for all $k\in \{1,2,\ldots, q\}$. 

Using this fact, the set of conditions (\ref{line0}--\ref{line2}) becomes 
\begin{align} 
a_k &\geq 0  \label{p1} \\
a_k &\leq  q-1,\label{p2} \\
a_k &\leq q-k,\label{p3}\\
a_1& \geq a_k \geq 1-k+a_1,\label{p4}
\end{align}
for all $k\in \{1,2,\ldots, q\}$.
Note that the condition $a_q =0$ has been dropped, since it follows from
(\ref{p1}) and (\ref{p3}).
Now we have a system of inequalities with only one index, and they can be combined together.
The lower bounds on $a_k$ are $0$  and $1-k+a_1$, hence this yields $a_k \geq \max\{1-k+a_1,0\}$.
The upper bounds are $q-1$, $q-k$, and $a_1$. The first of them is redundant, hence we have
$a_k \leq \min\{a_1,q-k\}$. All of this leads to a modified version of the previous proposition.
\begin{proposition}\label{mainprop}
Two-input $q$-state CA rule $f$ is number-conserving if and only if for all $i,j \in \mathcal{Q}$ 
\begin{equation*}
f(i,j)=i+a_{q-j}-a_{q-i},
\end{equation*}
where $a_1,a_2,\ldots, a_q$ is a sequence of integers satisfying
\begin{equation} \label{balanced}
\max\{1-k+a_1,0\} \leq a_k \leq \min\{a_1,q-k\}
\end{equation}
for all $k\in \{1,2,\ldots, q\}  $.
\end{proposition}

An integer sequence $a_1,a_2,\ldots,a_q$ satisfying (\ref{balanced}) will be called \emph{balanced
sequence}. Balanced sequences were introduced by Sheppard \cite{Sheppard76} as a representation
of \emph{properly labelled balanced graphs}, introduced earlier by Rosa~\cite{Rosa66}. 
Proposition \ref{mainprop} establishes one-to-one correspondence between balanced sequences and  number-conserving
CA rules with two inputs.
\section{Enumeration of number-conserving rules}
Enumeration of balanced sequences is a known combinatorial problem. In particular, one can show that
the number of balanced sequences with $q$ terms is equal to
$
2  \sum_{j=1}^{q/2} (j!)^2 j^{q-2j}
$
when $q$ is even, and 
$
2  \sum_{j=1}^{(q-1)/2} (j!)^2 j^{q-2j} +  
[(q-1)/2]! [(q+1)/2]!
$
 when $q$ is odd. The proof of this result can be found in  \cite{Sheppard76}, and will not be reproduced here.
Since the number of balanced sequences is equal to $\Lambda_{2,q}$, we immediately obtain the following theorem.
\begin{theorem}
There exist $\Lambda_{2,q}$ number-conserving CA rules with two inputs and $q$ states,
where

\begin{equation} \label{mainresult}
\Lambda_{2,q}=
\begin{cases}
\displaystyle 
2  \sum_{j=1}^{q/2} (j!)^2 j^{q-2j}  & \text{ if $q$ is even,}\\
\displaystyle 
2  \sum_{j=1}^{(q-1)/2} (j!)^2 j^{q-2j} +  \left(\frac{q+1}{2}\right)! \left(\frac{q-1}{2}\right)! 
\qquad &\text{ if $q$ is odd.} \
\end{cases}
\end{equation}
\end{theorem}
First ten terms of $\Lambda_{2,q}$ are
1, 2, 4, 10, 30, 106, 426, 1930, 9690, 53578. 
 Compared to  the sequence representing the total number of CA rules with $q$ states,  the sequence $\Lambda_{2,q}$ grows rather slowly.
 
We will now show that number-conserving CA rules are becoming increasingly rare as the number
of states increases. 
 Since the number of all two-input $q$-state CA rules is equal to $q^{q^2}$, for
 the case when $q$ is even one needs 
to consider the limit
\[ 
\lim_{q \to \infty}\frac{1}{q^{q^2}}\sum_{j=1}^{q/2} (j!)^2 j^{q-2j}.
\]
Using Stirling's approximation $(j!)^2 \approx 2 \pi j^{2j +1} e^{-2j}$, one obtains
\begin{align} \label{limq}
\lim_{q \to \infty}\frac{1}{q^{q^2}}\sum_{j=1}^{q/2} (j!)^2 j^{q-2j}=
\lim_{q \to \infty}\frac{2 \pi}{q^{q^2}}\sum_{j=1}^{q/2} j^{q+1} e^{-2j}.
\end{align}
Since 
\[
\frac{2 \pi}{q^{q^2}}\sum_{j=1}^{q/2} j^{q+1} e^{-2j} <
\frac{2 \pi}{q^{q^2}}\sum_{j=1}^{q/2} j^{q+1}  < \frac{2 \pi}{q^{q^2}} \frac{q}{2} 
\left(\frac{q}{2}\right)^{q+1} \
\xrightarrow{q\to \infty}0,  
\]
the limit (\ref{limq}) is equal to $0$. For the case of odd $q$ a very similar computation
yields the same result. This leads to the following corollary.
\begin{corollary}
The proportion of number-conserving rules among all CA rules with two inputs
tends to zero as $q$ increases, that is, 
$
\lim_{q \to \infty} \Lambda_{2,q} /q^{q^2}=0.
$
\end{corollary}

\section{Conclusions}
We have demonstrated that there exist one-to-one correspondence between number-conserving
two-input CA rules with $q$ states and balanced sequences of length $q$. 
The method of generating and counting all balanced sequences of length $q$ has been described
in  \cite{Sheppard76}. Therefore, our result completely solves the enumeration problem for
two-input number-conserving CA rules with $q$ states.

It should be possible to generalize this result to rules with larger neighbourhood sizes,
by observing that eq. (\ref{NScond}) for $x_1=0$ becomes 
\begin{align} 
f(0,x_2,\ldots,x_n) =  \sum_{k=1}^{n-1}\big(
&f(\underbrace{0,0,\ldots,0}_k,x_2,x_3,\ldots,x_{n-k+1})\notag\\
-&f(\underbrace{0,0,\ldots,0}_{k+1},x_2,\ldots,x_{n-k})\big),
\end{align}
which further simplifies to $f(0,x_2,\ldots,x_n) =f(0,x_2,\ldots,x_n)$. This means that 
$f(0,x_2,\ldots,x_n)$  for $x_2,\ldots,x_n\in \mathcal{Q}$  can again be treated 
as free parameters, and that an equivalent of Proposition~\ref{th2} can be constructed. It is not
immediately clear, however, how to count sequences satisfying the resulting set of inequalities.

Another interesting problem is the connection with graph theory. Balanced sequences
were originally proposed to represent properly labelled balanced graphs. This means that 
each number-conserving CA with two inputs has a natural representation
as a labelled graph. Graph representation can be useful to explore symmetries
of number-conserving rules as well as some features of their dynamics. This
problem is currently under investigation and will be reported elsewhere.
\section{Acknowledgments}
One of the authors (H.F.) acknowledges financial support from NSERC (Natural Sciences and Engineering Research Council of Canada) in the form of the Discovery Grant.


\end{document}